\documentclass[prd,superscriptaddress,a4paper,showpacs,showkeys,nofootinbib]{revtex4}
\usepackage{graphicx}
\usepackage{graphics}
\usepackage{epsfig}
\usepackage{dcolumn}
\usepackage{bm}
\usepackage{multirow}
\usepackage{tabularx}
\usepackage{hyperref}
\usepackage{commath}
\usepackage{epstopdf}
\usepackage[T1]{fontenc}
\usepackage{geometry}
\usepackage{color}

\def\be{\begin{equation}}
\def\ee{\end{equation}}

\providecommand{\ee}{e$^+$e$^-$}

\makeatother

\begin{document}
%
%
\title{Exclusive axionlike particle production by gluon -- induced interactions \\ in hadronic collisions}


\author{V. P. Gon\c calves}

\email[]{barros@ufpel.edu.br}

\affiliation{Instituto de F\'{\i}sica e Matem\'atica, Universidade Federal de
Pelotas (UFPel),\\
Caixa Postal 354, CEP 96010-090, Pelotas, RS, Brazil}

\author{W. K. Sauter}

\email[]{werner.sauter@ufpel.edu.br}

\affiliation{Instituto de F\'{\i}sica e Matem\'atica, Universidade Federal de
Pelotas (UFPel),\\
Caixa Postal 354, CEP 96010-090, Pelotas, RS, Brazil}



\begin{abstract}
The exclusive production of axionlike particles (ALPs) by gluon -- induced interactions    is investigated in this exploratory study considering  $pp$ and $PbPb$ collisions for the energies of the next run of the Large Hadron Collider (LHC).  Assuming the Duhram model, we estimate the associated cross sections for the rapidity ranges probed by central and forward detectors. A comparison with the predictions for the exclusive ALP production by photon -- induced interactions is presented. Our results indicate that the contribution of gluon -- induced interactions is nonnegligible and can become dominant in $pp$ collisions for small values of the ALP mass.
\end{abstract}




\maketitle

\begin{figure}[t]
\begin{tabular}{cc}
\hspace{-1cm}
{\psfig{figure=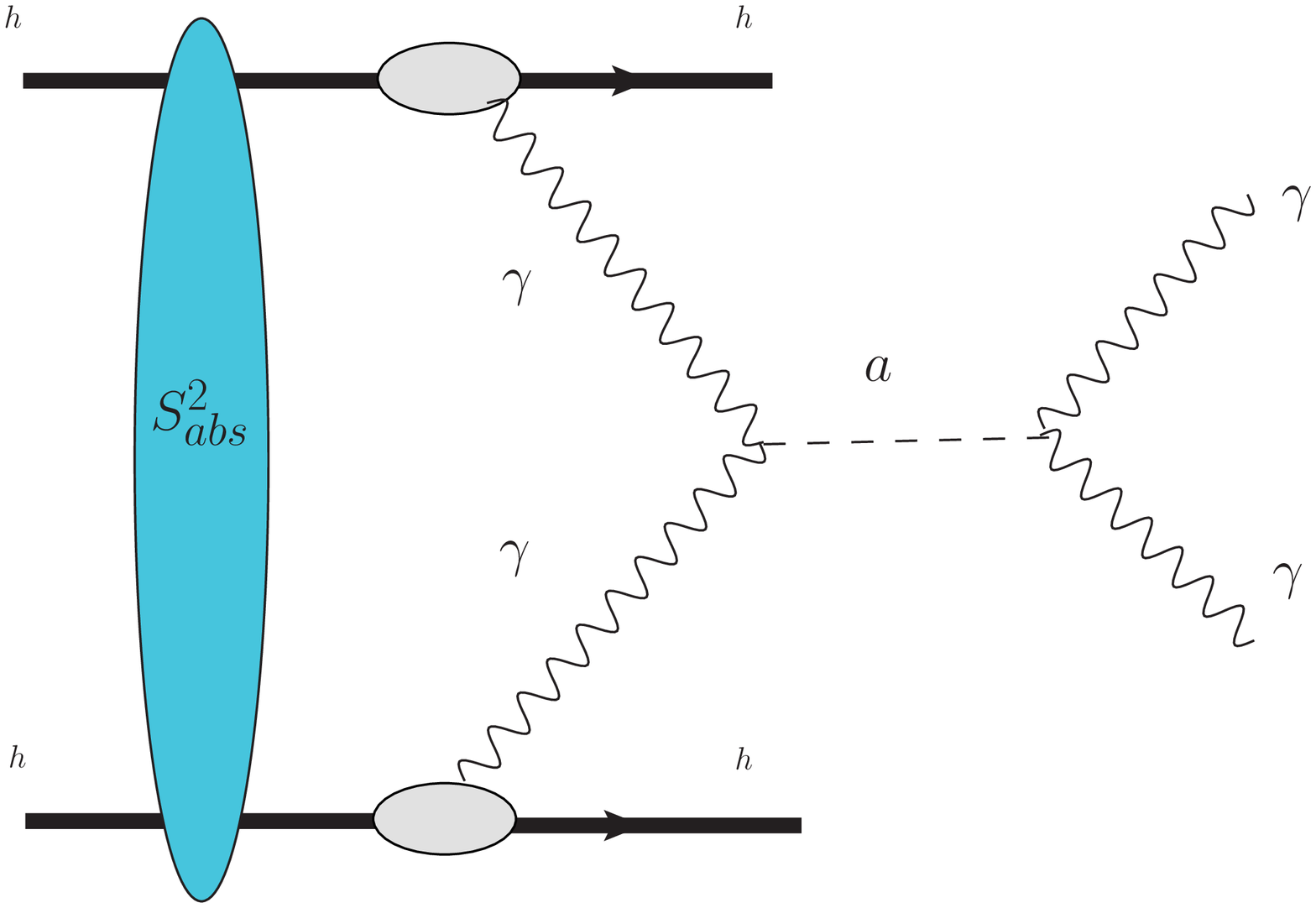,width=8.0cm}} &
{\psfig{figure=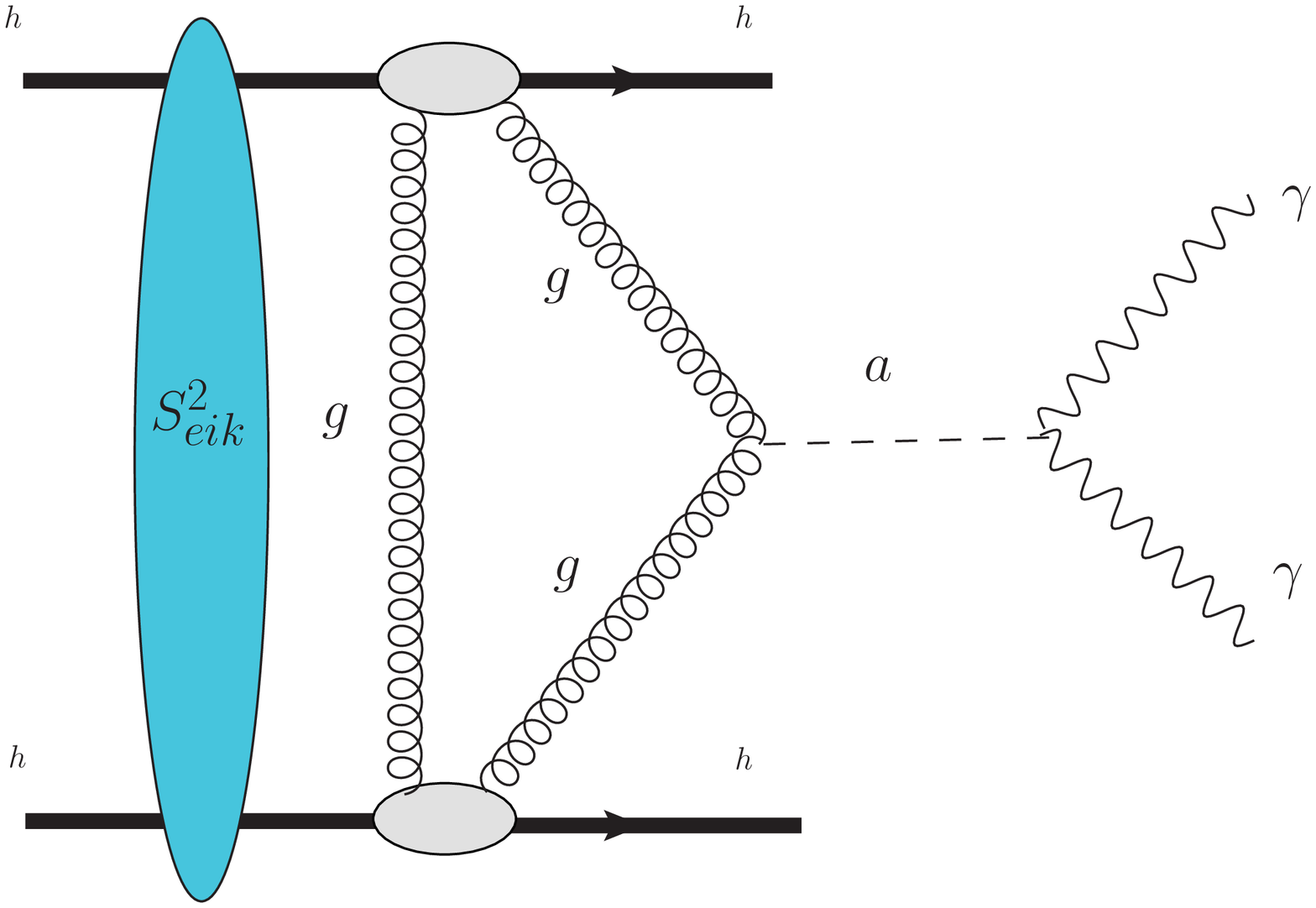,width=8.0cm}}  \\
(a) & (b) \\ 
\end{tabular}                                                                                                                       
\caption{Diphoton production in hadronic collisions by  the (a)   $\gamma \gamma \rightarrow a \rightarrow\gamma \gamma$  and (b) $gg \rightarrow a \rightarrow\gamma \gamma$ subprocesses.}
\label{fig:diagram}
\end{figure}

Axionlike particles (ALPs) are pseudo -- Nambu -- Goldstone bosons  predicted to occur in many extensions of the Standard Model (SM) due to the spontaneous breaking of a global symmetry and are candidates to constitute the cosmological dark matter. Such particles are expected to be characterized by a small mass $m_a$ and by  suppressed couplings to the SM particles. During recent years, several authors have proposed the searching for axionlike particles in $e^+e^-$, $ep$, $\nu p$, $pp$, $pA$ and $AA$ collisions as well in laser beam experiments 
(See e.g. Refs. \cite{Jaeckel:2015jla,Bauer:2017ris,knapen,Aloni:2018vki,royon,Aloni:2019ruo,Bauer:2018uxu,Yue:2019gbh,Ebadi:2019gij,Alves:2019xpc}). One of the more promissing alternatives is the searching for ALPs  in ultrarelativistic heavy ion collisions, considering a diphoton system as the final state, as represented in Fig. \ref{fig:diagram} (a). In these collisions, the  photon -- photon luminosity that scales with $Z^4$, where $Z$ is number of protons in the nucleus, which implies a large enhancement of the ALP production cross section. Moreover, the resulting  final state is very clean, consisting  of the diphoton system,  two intact nuclei and  two rapidity gaps, i.e. empty regions  in pseudo-rapidity that separate the intact very forward nuclei from the $\gamma \gamma$ system. As recently demonstrated in Ref. \cite{rafael_axion}, the backgrounds associated to the Light - by - Light (LbL) scattering and to the diffractive diphoton production can be strongly reduced in $PbPb$ collisions by the exclusivity cuts and that a forward detector, as the LHCb, is ideal to probe an ALP with small mass. Similar study for $pp$ collisions using the proton tagging technique was performed in Ref. \cite{royon1}, which demonstrated that such collisions can constrain ALPs masses in the range $0.5 \le m_a \le 2.0$ TeV. All these studies focused in the exclusive diphoton production by  photon -- induced interactions, characterized by the $\gamma \gamma \rightarrow a \rightarrow \gamma \gamma$ subprocess. 
Our goal in this letter is to complement these previous studies, by estimating, for the first time, the exclusive diphoton production in gluon -- induced interactions, where the elementary subprocess is the $gg \rightarrow a \rightarrow \gamma \gamma$ reaction. Such process is represented in Fig. \ref{fig:diagram} (b). As the gluons exchanged between the incident particles  (protons or nuclei) are in a color singlet configuration, the final state  will also be characterized by two rapidity gaps and two intact hadrons. Therefore, such process is an irreducible background for the searching of ALPs in photon -- induced interactions. In reality, as we will shown below, the exclusive ALP production in gluon -- induced interactions can become dominant in $pp$ collisions for ALPs masses smaller than 100 GeV. In this exploratory study, we will present predictions for  fixed values for the ALP couplings to gluons and to photons. We postpone for a future publication a more detailed analysis \cite{nos_fpmc}. Currently, we are implementing the gluon -- induced reaction in the Forward Physics Monte Carlo (FPMC) \cite{fpmc}, which will allow us to derive the expected excluded regions of the parameter space of the ALP model taking into account of the exclusivity cuts.

Initially, we present a brief review of formalism used to estimate the exclusive ALP production by gluon -- induced interactions in $pp$ and $PbPb$ collisions. Such process, represented in Fig. \ref{fig:diagram} (b),  will be described using the Durham model \cite{kmr_prosp}, proposed many years ago and extensively discussed in the literature (For a review see, e.g. Ref. \cite{review_kmr}). In this model, the ALP production occurs by the hard subprocess $gg \rightarrow a$ and a second $t$ -- channel gluon is needed to screen the colour flow across the rapidity gap intervals. The decay of the ALP into two photons is described by the branching ratio $\mbox{BR}(a \rightarrow \gamma \gamma) $. As a consequence, the  total cross section   can be expressed  as follows
\begin{equation}
 \sigma \left(h h \rightarrow h \otimes \gamma \gamma \otimes h;s \right)  = \int dy \, \int \frac{dW^2}{W^2} \langle {\mathcal S}_{eik}^2 \rangle \, {\mathcal L}_{excl} \, \hat{\sigma}(gg \rightarrow a)  \,\,\mbox{BR}(a \rightarrow \gamma \gamma)     \,\,, 
\label{eq:kmr}
\end{equation}
where  $\sqrt{s}$ is center - of - mass energy of the hadronic collision, $\otimes$ characterizes a rapidity gap in the final state, $y$ the rapidity of the final state, $W$ is the invariant mass of the final state and $\langle {\mathcal S}_{eik}^2 \rangle$ is the gap survival probability. Moreover, $\hat{\sigma}(gg \rightarrow a)$ is the cross section associated to the $gg \rightarrow a$ subprocess, which is given by \cite{kmr_prosp}
\begin{equation} 
\hat{\sigma}_{gg \rightarrow a} = \frac{2\pi^2}{m_a} \Gamma({a \rightarrow gg}) \times \delta(W^2 - m_a^2),
\end{equation}
where $\Gamma (a \rightarrow gg)$ stand for the partial decay width of the ALP in a pair of gluons. The quantity ${\mathcal L}_{excl}$ is the effective luminosity for exclusive processes, defined by
\begin{equation}
 {\mathcal L}_{excl} = \left[ {\cal{C}} \int \frac{dQ_t^2}{Q_t^4} f_g(x_1,x_1^{\prime},Q_t^2, \mu^2) f_g(x_2,x_2^{\prime},Q_t^2, \mu^2)\right]^2\,\,,
 \label{eq:lum}
\end{equation}
where ${\cal{C}} = \pi/[(N_c^2 - 1)b]$, with $b$ being the $t$-slope ($b = 4$ GeV$^{-2}$ in what follows), $Q_t^2$ is the virtuality of the soft gluon needed for color  screening, $x_1$ and $x_2$   the longitudinal momentum of the gluons which participate of the hard subprocess  and $x_1^{\prime}$ and $x_2^{\prime}$ the longitudinal momenta of the spectator gluon.
 The quantities $f_g$ are the 
skewed unintegrated gluon distributions $f_g$. At leading logarithmic approximation, it is possible to express $f_g(x,x^{\prime},Q_t^2, \mu^2)$ in terms of the conventional integral gluon density $g(x)$ and the Sudakov factor $T$, which ensures that the active gluons that participate of the hard process do not radiate in the evolution from $Q_t$ up to the hard scale $\mu \approx m_a/2$.
In this letter we will calculate  $f_g$ in the proton case considering that  the integrated gluon distribution $xg(x,Q_T^2)$ is described by the  MMHT2014 parametrization \cite{mmht}. In the nuclear case we will include the shadowing effects in $f_g^A$ considering that the nuclear gluon distribution is given by the  nCTEQ parametrization~\cite{ncteq}.  In order to obtain a realistic prediction for the exclusive ALP production, we need to take into account of the  soft interactions that are expected to lead to extra production of particles, which will  destroy the rapidity gaps in the final state and modify the associated cross sections \cite{bjorken}. In the Durham model, such soft corrections are included in the  eikonal factor $\langle {\mathcal S}_{eik}^2 \rangle$. In our study we will assume that the hard process occurs on a short enough timescale such that the physics that generate the additional particles can be factorized and estimated using an soft approach for hadronic interactions constrained by the  diffractive data. Following Ref.  \cite{kmr_prosp},  we will assume that $\langle {\mathcal S}^2_\mathrm{excl} \rangle = 3 \, \%$ for $pp$ collisions at the LHC  energy. The value of the survival probability for nuclear collisions  is still an open question. In what follows, we will consider the conservative estimate  proposed in Ref.  \cite{radion}, in which  $\langle {\mathcal S}^2_\mathrm{excl} \rangle_{A_1A_2} = \langle {\mathcal S}^2_\mathrm{excl} \rangle_{pp}/(A_1 \cdot A_2)$. However, it is important to emphasize that smaller values were derived in Refs. ~\cite{miller,anderson} using the Glauber model. Consequently, our predictions for the nuclear case can be considered an upper bound for the cross sections.

For completeness of our study, the exclusive ALP production by photon -- induced interactions will also be estimated. For the process represented in Fig. \ref{fig:diagram} (a), the total cross section can be expressed as follows \cite{upc1,epa}:
 \begin{eqnarray}
\sigma \left(h h \rightarrow h \otimes \gamma \gamma \otimes h;s \right)   
&=& \int \mbox{d}^{2} {\mathbf r_{1}}
\mbox{d}^{2} {\mathbf r_{2}}  
\mbox{d}y \mbox{d}W  \frac{W}{2}  \,   N\left(\omega_{1},{\mathbf r_{1}}  \right)
 N\left(\omega_{2},{\mathbf r_{2}}  \right ) S^2_{abs}({\mathbf b})\,\hat{\sigma}(\gamma \gamma \rightarrow a) \, \mbox{BR}(a \rightarrow \gamma \gamma)  
  \,\,,
\label{cross-sec-2}
\end{eqnarray}
 where  $N(\omega_i, {\mathbf r}_i)$ is the equivalent photon spectrum, which allows to estimate the number the photons  with energy $\omega_i$ at a transverse distance ${\mathbf r}_i$  from the center of hadron, defined in the plane transverse to the trajectory. Moreover, the invariant mass is given by $W = \sqrt{4 \omega_1 \omega_2}$. The cross section for  the $\gamma \gamma \rightarrow a$ subprocess is given by \cite{Low}
\begin{equation} \label{eq:epa}
\hat{\sigma}_{\gamma \gamma \rightarrow a} = \frac{8\pi^2}{m_a} \Gamma({a \rightarrow \gamma \gamma}) \times \delta(W^2 - m_a^2),
\end{equation}
where $\Gamma (a \rightarrow \gamma \gamma)$ is the partial decay width of the ALP into a pair of photons.  
 The 
 absorptive factor $S^2_{abs}({\mathbf b})$, which depends on the impact parameter ${\mathbf b}$ of the hadronic collision, insure the dominance of the electromagnetic interaction by excluding the overlap between the colliding hadrons. 
In our calculations, we will estimate the photon spectrum assuming a pointlike form factor for the hadron.
Moreover, the absorptive factor will be estimated using the model proposed by 
Baur and Ferreira - Filho \cite{Baur_Ferreira}, where $S^2_{abs}({\mathbf b}) = \Theta\left(
\left|{\mathbf b}\right| - 2 R
 \right )  = 
\Theta\left(
\left|{\mathbf r_{1}} - {\mathbf r_{2}}  \right| - 2 R
 \right )$
where $R$ is the hadron radius, being equal to 0.7 fm for a proton and 1.2 $A^{1/3}$ fm for a nuclei. Such model treats the hadrons as hard spheres with radius $R$ and assumes that the probability to have a hadronic interaction when $b > 2 R$ is zero. It is important to emphasize that other models can be used to treat the photon flux and the absorptive factor, as discussed in detail in Ref. \cite{celsina}. However, for the values of the ALP masses considered in this letter, the predictions from these different approaches are almost identical.

 \begin{figure}
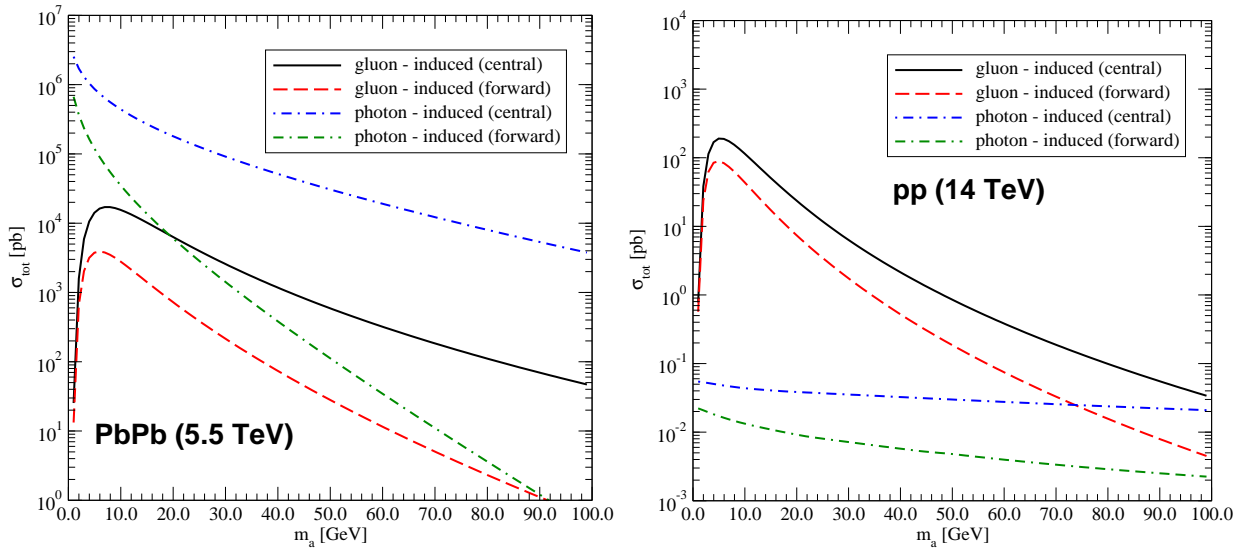

\begin{tabular}{cc}
\hspace{-1cm}
{\psfig{figure=pbpb-axion2.eps,width=8.0cm}} &
{\psfig{figure=pp-axion.eps,width=8.0cm}}  
\end{tabular}                                                                                                                       
\caption{ALP mass dependence of the cross sections for the diphoton production in  $PbPb$ (left panel) and $pp$ (right panel) collisions.}
\label{fig:resultados}
\end{figure}

The main input in our calculations are the ALP mass $m_a$ and the partial decay widths $\Gamma (a \rightarrow gg)$ and $\Gamma (a \rightarrow \gamma \gamma)$. At leading order, the decay rates of an ALP into two gluons and two photons can be written as (See e.g. Ref. \cite{ebadi})
\begin{eqnarray}
\Gamma(a \rightarrow gg) = 8 \times \left(\frac{c_{gg}}{f_a}\right)^2 \frac{m_a^3}{4 \pi}
\end{eqnarray}
and
\begin{eqnarray}
\Gamma(a \rightarrow \gamma \gamma) =  \left(\frac{c_{\gamma \gamma}}{f_a}\right)^2 \frac{m_a^3}{4 \pi}\,\,.
\end{eqnarray}
In what follows we will estimate the cross sections for the ALP production in gluon and photon -- induced interactions as a function of the ALP mass $m_a$ considering $pp$ collisions at $\sqrt{s} = 14$ TeV and $PbPb$ collisions at $\sqrt{s} = 5.5$ TeV, which are the energies of the next run of the LHC. We will present predictions considering the typical rapidity ranges probed by  central detectors  ($|y| \le 2.0$), as e.g. by the ALICE, ATLAS and CMS detectors, as well as by a forward detector ($ 2.0 \le y \le 4.5$), as the LHCb one.  In order to illustrate how important can be the gluon -- induced process, we will assume in this exploratory study that $c_{gg}/f_a = c_{\gamma \gamma}/f_a = 10^{-4}$ GeV$^{-1}$.
Moreover, as in previous studies \cite{knapen,royon,rafael_axion} we will estimate the cross sections under the assumption that $\mbox{BR}(a \rightarrow \gamma \gamma) = 1$ and, therefore, our results are a upper limit for the cross section. As pointed out before, a detailed analysis of the ALP production considering the full  parameter space ($m_a,c_{gg}/f_a$), obtained using  the FPMC event generator  and taking into account of the exclusivity cuts, will be presented in  a future publication \cite{nos_fpmc}.  

\begin{figure}
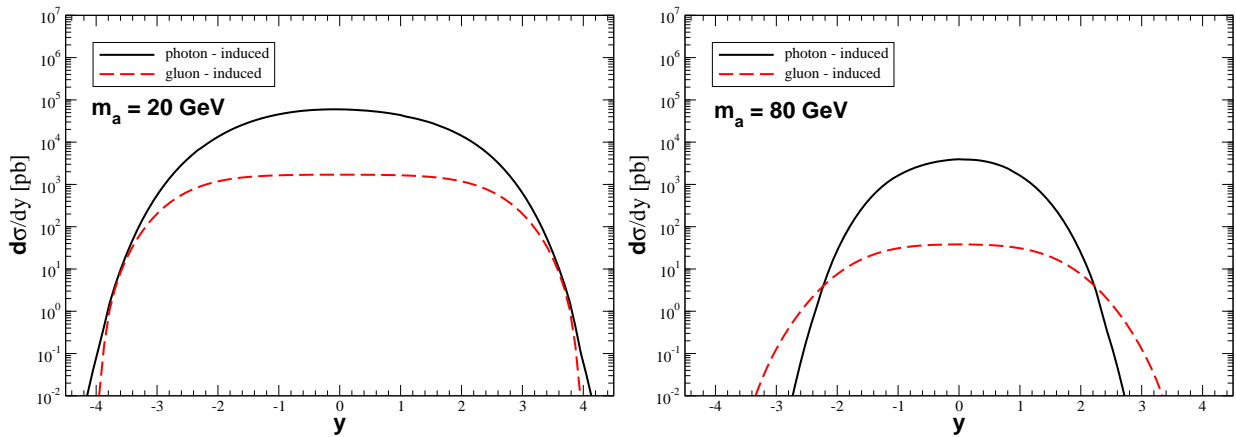

\begin{tabular}{cc}
\hspace{-1cm}
{\psfig{figure=dsdy_20.eps,width=8.0cm}} &
{\psfig{figure=dsdy_80.eps,width=8.0cm}}  
\end{tabular}                                                                                                                       
\caption{Rapidity distributions for the diphoton production in $PbPb$ collisions considering two distinct values of the ALP mass.}
\label{fig:rap}
\end{figure}

Our predictions  are presented in Fig. \ref{fig:resultados}. For $PbPb$ collisions  (left panel), the ALP production by photon -- induced interactions is dominant for a large range of ALP masses. Such behaviour is directly associated to the $Z^4$ -- enhancement of the cross section in $\gamma \gamma$ interactions. Our results indicate that for a central detector, the gluon -- induced contribution is almost two orders of magnitude smaller than the photon -- induced one for the mass range considered. On the other hand,   the gluon --  and photon -- induced interactions become similar for large values of $m_a$ in the kinematical range probed by a forward detector. Such behavior is associated with the fact that the rapidity 
distribution associated with photon -- induced interaction becomes more 
narrow in rapidity  than the 
gluon -- induced prediction for larger values of the ALP masses, as demonstrated in Fig. \ref{fig:rap} where we 
present our predictions for the rapidity distributions considering
 two values of $m_a$.

As discussed in Refs. \cite{rafael_axion,royon1,Goncalves:2020vvw}, the diphotons produced in photon -- induced interactions are expected to be characterized by a smaller pair transverse momentum $p_T^{\gamma \gamma}$ than those produced by the Durham process. In addition, the acoplanarity ($A_{CO}$) of the photon pair is predicted to be significantly broader in gluon -- induced interactions. Another important aspect  in gluon -- induced interactions at heavy ion collisions, is that the nucleus is expected to dissociates due to the gluons recoil, which results in the emission of neutrons that are detectable in the zero -- degree calorimeters (ZDCs) located along the beam axis. In principle, such characteristics can be used to suppress the photon -- induced processes in $PbPb$ collisions. In particular, the analysis performed in Ref. \cite{Aaboud:2017bwk} have demonstrated that the observed events with a ZDC signal, corresponding to at least one neutron, agree with the  Durham predictions normalized to the data associated to events that satisfy $A_{CO} > 0.02$. It is important to emphasize that although the normalization of the Durham predictions has a large theoretical uncertainty associated to the distinct treatments for the  survival probability for nuclear collisions, that is not a limitation for the ALP search.   
As the soft corrections for the ALP production  are predicted to be the same than those present in the continuum diphoton production by gluon induced interactions,  we can fix the normalization using the data for the   smooth diphoton invariant mass distribution associated to the continuum  and search for the resonant bump  generated by the ALP decaying into two photons. Such procedure can be applied for high and low diphoton masses ($M_{\gamma \gamma}$), but for low values of $M_{\gamma \gamma}$ ($\le 5.0$ GeV) it is important to take into account the background associated to the resonances that can be produced by gluon -- gluon interactions and that decay into two photons as e.g. $\eta, \, \eta_c$ and $\chi_c$. A detailed analysis of this background will be presented in Ref. 
\cite{nos_fpmc} following the study performed in  Ref. \cite{mariola}.


In Fig. \ref{fig:resultados} (right panel) we present our predictions for $pp$ collisions. One has that for $m_a \lesssim 70$ GeV, the gluon -- induced interactions dominate the exclusive ALP production, being $\approx 3$ orders of magnitude larger than the photon -- induced one for $m_a \approx 5$ GeV.
However, both contributions become similar for $m_a \approx 100$ GeV and the photon -- induced interactions dominate for larger masses.  The dominance of the  gluon -- induced interactions for the production of ALPs with small mass is directly associated to the large number of gluons in the incident protons at small values of the Bjorken - $x$ variable. At high energies and small mass, the main contribution comes from small - $x$, which implies a very large effective luminosity [See Eq. (\ref{eq:lum})]. The decreasing of the cross section with the ALP mass is associated to the rapid decreasing of the gluon distribution at large values of $x$. In contrast, the decreasing of the effective $\gamma \gamma$ luminosity is slowly in comparison to $ {\mathcal L}_{excl} $ due to behaviour of the equivalent photon flux with the photon energy, which is almost flat in the energy range considered. 

Our results indicate that for the mass range that can be probed in $pp$ collisions using the proton tagging detectors  ($m_a \ge 200$ GeV) \cite{afp1,afp2,pps}, the gluon -- induced contribution is small for the coupling considered. In principle, the separation of the gluon -- induced processes can be performed by imposing cuts in the pair transverse momentum  and acoplanarity, as discussed for nuclear collisions. However, differently from $PbPb$ collisions, one of the main challenges present in the study of central exclusive processes at $pp$ collisions is the experimental separation of these events, especially during the high pile-up running. Pile-up is referred to the multiple soft proton-proton interactions in each bunch crossing of the LHC which  occur simultaneously with the hard process associated to the primary vertex of interest. The average number of pile-up interactions per bunch crossing  during the  Run I to II varies between 20-50 and the expectation for high luminosity LHC is in the range 140-200. As a consequence, the signal events associated to exclusive processes suffer from the background that arise from the other processes with  common final state particles and from inclusive processes which are coincided with the pile-up protons. The presence of forward detectors with high resolution on
momentum and arrival time of protons have been used to suppress background contributions \cite{afp1,afp2,pps}. In principle, the measurement of the forward protons permits to predict the kinematics of centrally produced state, which can be measured separately. The matching between these two measurements can lead to several orders of magnitude suppression in the inclusive background processes.
In addition, the high correlation between the primary vertex
 displacement in the $z$-direction and the arrival time of both tagged protons to the timing forward detectors  present in  exclusive processes, can also be used to reduce the inclusive background contribution. Such reduction depends on  the
timing resolution of time of flight detectors. The results presented in Ref. \cite{royon} for the search of axion-like particles with proton tagging at the LHC, produced by photon -- induced interactions, indicated that the suppression of the pile - up is possible. A similar result is expected for gluon -- induced interactions, but a more definitive conclusion deserves a detailed study \cite{nos_fpmc}.

The kinematical range of small diphoton invariant masses, where the  ALP production by gluon -- induced interactions become dominant, can be probed at the LHC  in the case of special runs with a modified  optics, which will be characterized by a low pile - up. One shortcoming  is  the reduced luminosity expected for these runs, which will be ${\cal{O}}(10 \,$pb$^{-1})$. As a consequence, we expect a very small number of events associated to ALP with masses  larger than 20 GeV. On the other hand, for $m_a \approx 10$ GeV and $c_{gg}/f_a  = 10^{-4}$ GeV$^{-1}$, we predict $\approx 10^3$ events. Such rate increases by a factor 2 for $m_a \approx 5$ GeV. However, for this small value of mass,  the  backgrounds associated to the production of other resonances, the misidentification of low - $p_T$ photons and from diphotons that arise from the  photoproduced $\pi^0$ pairs, are expected to strongly increase, which becomes the separation of these events a hard task. A more detailed analysis will be presented in Ref. \cite{nos_fpmc}. An alternative is the increasing of the integrated luminosity in the special runs for values of the order of 0.1 fb$^{-1}$, which can be reached in a week of data taking. Such configuration will allow to investigate in detail the ALP production by gluon -- induced interactions at $pp$ collisions.


As a summary, in this letter we have performed, for the first time, an exploratory study of  the exclusive ALP production by gluon -- induced interactions in hadronic collisions. Our results indicate that the contribution of such process can be nonnegligible in $pp$ collisions, which motivates  a more detailed analysis to determine if this channel can be used to searching for the axionlike particles and to constrain its main properties.

\begin{acknowledgments}
VPG acknowledge very useful discussions about ALP production in photon - induced interactions with D. E. Martins and M. S. Rangel.
This work was  partially financed by the Brazilian funding
agencies CNPq,   FAPERGS and INCT-FNA (processes number 
464898/2014-5).
\end{acknowledgments}


\begin{thebibliography}{99}

\bibitem{Jaeckel:2015jla} 
  J.~Jaeckel and M.~Spannowsky,
  Phys.\ Lett.\ B {\bf 753}, 482 (2016).

\bibitem{Bauer:2017ris} 
  M.~Bauer, M.~Neubert and A.~Thamm,
  JHEP {\bf 1712}, 044 (2017).


\bibitem{knapen} 
  S.~Knapen, T.~Lin, H.~K.~Lou and T.~Melia,
  Phys.\ Rev.\ Lett.\  {\bf 118}, no. 17, 171801 (2017).


\bibitem{Aloni:2018vki} 
  D.~Aloni, Y.~Soreq and M.~Williams,
  Phys.\ Rev.\ Lett.\  {\bf 123}, no. 3, 031803 (2019).

\bibitem{royon} 
  C.~Baldenegro, S.~Hassani, C.~Royon and L.~Schoeffel,
  Phys.\ Lett.\ B {\bf 795}, 339 (2019).



\bibitem{Aloni:2019ruo} 
  D.~Aloni, C.~Fanelli, Y.~Soreq and M.~Williams,
  Phys.\ Rev.\ Lett.\  {\bf 123}, no. 7, 071801 (2019).
  
\bibitem{Bauer:2018uxu} 
  M.~Bauer, M.~Heiles, M.~Neubert and A.~Thamm,
  Eur.\ Phys.\ J.\ C {\bf 79}, no. 1, 74 (2019).  
  
  \bibitem{Yue:2019gbh} 
  C.~X.~Yue, M.~Z.~Liu and Y.~C.~Guo,
  Phys.\ Rev.\ D {\bf 100}, no. 1, 015020 (2019).
  
  \bibitem{Ebadi:2019gij} 
  J.~Ebadi, S.~Khatibi and M.~Mohammadi Najafabadi,
  Phys.\ Rev.\ D {\bf 100}, no. 1, 015016 (2019).
  
  
  
  \bibitem{Alves:2019xpc}   
A.~Alves, A.~G.~Dias and D.~D.~Lopes,
JHEP \textbf{08}, 074 (2020)
  
\bibitem{rafael_axion}
R.~Coelho, V. P. ~Goncalves, D.~Martins and M.~Rangel,
Phys. Lett. B \textbf{806}, 135512 (2020)  
  
  
  \bibitem{royon1}
C.~Baldenegro, S.~Fichet, G.~von Gersdorff and C.~Royon,
JHEP \textbf{06}, 131 (2018)
  
  \bibitem{nos_fpmc}
 V. P. ~Goncalves, D.~Martins,  M.~Rangel and W. K. Sauter. {\it work in progress.} 
  
\bibitem{fpmc} 
  M.~Boonekamp, A.~Dechambre, V.~Juranek, O.~Kepka, M.~Rangel, C.~Royon and R.~Staszewski,
  arXiv:1102.2531 [hep-ph].



\bibitem{kmr_prosp}
  V.~A.~Khoze, A.~D.~Martin and M.~G.~Ryskin,
  Eur.\ Phys.\ J.\  C {\bf 23}, 311 (2002);  Eur.\ Phys.\ J.\  C {\bf 24}, 581 (2002).
  
\bibitem{review_kmr}
L.~Harland-Lang, V.~Khoze and M.~Ryskin,
Int. J. Mod. Phys. A \textbf{29}, 1446004 (2014)  



\bibitem{mmht} 
  L.~A.~Harland-Lang, A.~D.~Martin, P.~Motylinski and R.~S.~Thorne,
  Eur.\ Phys.\ J.\ C {\bf 75}, no. 5, 204 (2015)
  
\bibitem{ncteq}
K.~Kovarik, A.~Kusina, T.~Jezo, D.~Clark, C.~Keppel, F.~Lyonnet, J.~Morfin, F.~Olness, J.~Owens, I.~Schienbein and J.~Yu,
Phys. Rev. D \textbf{93}, no.8, 085037 (2016)





\bibitem{bjorken} 
  J.~D.~Bjorken,
  Phys.\ Rev.\ D {\bf 47}, 101 (1993).




\bibitem{radion}
  V.~P.~Goncalves and W.~K.~Sauter,
  Phys.\ Rev.\ D {\bf 82}, 056009 (2010).



\bibitem{miller}
  E.~Levin and J.~Miller,
  arXiv:0801.3593 [hep-ph].

\bibitem{anderson}
E.~Basso,~V.~P.~Goncalves, A.~K.~Kohara, M.~S.~Rangel, Eur. Phys. J. C  {\bf 77}, 600 (2017).



\bibitem{epa} 
  V.~M.~Budnev, I.~F.~Ginzburg, G.~V.~Meledin and V.~G.~Serbo,
  Phys.\ Rept.\  {\bf 15}, 181 (1975).
  

\bibitem{upc1}
C. A. Bertulani and G. Baur, { Phys. Rep.} {\bf 163}, 299 (1988).


\bibitem{Low}       F.~E.~Low,  Phys.\ Rev.\  {\bf 120}, 582 (1960). 	


\bibitem{Baur_Ferreira} 
  G.~Baur and L.~G.~Ferreira Filho,
  Nucl.\ Phys.\ A {\bf 518}, 786 (1990).



 \bibitem{celsina} 
  C.~Azevedo, V.~P.~Goncalves and B.~D.~Moreira,
  Eur.\ Phys.\ J.\ C {\bf 79}, no. 5, 432 (2019). 



\bibitem{ebadi}
J.~Ebadi, S.~Khatibi and M.~Mohammadi Najafabadi,
Phys. Rev. D \textbf{100}, no.1, 015016 (2019)
  
  
\bibitem{Goncalves:2020vvw}
V.~P.~Goncalves, D.~E.~Martins and M.~S.~Rangel,
Eur. Phys. J. C \textbf{80}, no.9, 841 (2020)


\bibitem{Aaboud:2017bwk}
M.~Aaboud \textit{et al.} [ATLAS],
Nature Phys. \textbf{13}, no.9, 852-858 (2017)  


\bibitem{mariola}
M.~Klusek-Gawenda, R.~McNulty, R.~Schicker and A.~Szczurek,
Phys. Rev. D \textbf{99}, no.9, 093013 (2019)



\bibitem{afp1}
  L.~Adamczyk {\it et al.},
 ``Technical Design Report for the ATLAS Forward Proton Detector,''
  CERN-LHCC-2015-009, ATLAS-TDR-024.

\bibitem{afp2}
  M.~Tasevsky [ATLAS Collaboration],
 ``Status of the AFP project in the ATLAS experiment,''
  AIP Conf.\ Proc.\  {\bf 1654}, 090001 (2015).
  
\bibitem{pps}
  M.~Albrow {\it et al.} [CMS and TOTEM Collaborations],
  ``CMS-TOTEM Precision Proton Spectrometer,''
  CERN-LHCC-2014-021, TOTEM-TDR-003, CMS-TDR-13.
  
    

\end{thebibliography}
 \end{document}